\documentclass[preprint,showpacs]{revtex4}
\usepackage{graphicx}
\usepackage{dcolumn}
\usepackage{bm}
\begin{document}
\title{Electronic structure of Pr$_{0.67}$Ca$_{0.33}$MnO$_3$ near the Fermi level studied by ultraviolet photoelectron and x-ray absorption spectroscopy}
\author{M. K. Dalai, P. Pal, and B. R. Sekhar}
\affiliation{Institute of Physics, Sachivalaya Marg, Bhubaneswar-751005, India}
\author{N. L. Saini}
\affiliation{Dipartimento di Fisica, Universita' di Roma ``La Sapienza", 
             Piozzale Aldo Moro 2, 00185 Roma, Italy}
\author{R. K. Singhal and K. B. Garg}
\affiliation{Department of Physics, University of Rajasthan, 
Jaipur-302004, India}
\author{B. Doyle}
\affiliation{Laboratorio Nazionale TASC, INFM-CNR, S.S.14, km 163.5, Area 
Science Park, 34012 Basovizza (TS), Italy}
\author{S. Nannarone}
\affiliation{Laboratorio Nazionale TASC, INFM-CNR, S.S.14, km 163.5, Area 
Science Park, 34012 Basovizza (TS), Italy \\
and Dipartimento di Ingegneria dei Materiali e dell'Ambiente, Universit$\grave{a}$ di Modena e Reggio Emilia, Via Vignolese 905, 41100 Modena, Italy}
\author{C. Martin and F. Studer}
\affiliation{Laboratoire CRISMAT, UMR 6508, ISMRA, Boulevard du Marechal Juin, 14050 Caen, France}

\begin{abstract}

We have investigated the temperature-dependent changes in the near-$E$$_F$ 
occupied and unoccupied states of Pr$_{0.67}$Ca$_{0.33}$MnO$_3$ which 
shows the presence of ferromagnetic and antiferromagnetic phases. The 
temperature-dependent changes in the charge and orbital degrees of freedom 
and associated changes in the Mn 3$d$ - O 2$p$ hybridization result in varied 
O 2$p$ contributions to the valence band. A quantitative estimate of the 
charge transfer energy ($E$$_{CT}$) shows a larger value compared to the 
earlier reported estimates. The charge localization causing the large 
$E$$_{CT}$ is discussed in terms of different models including the 
electronic phase separation.

\end{abstract}
\maketitle
\section{INTRODUCTION}

Recently, a lot of attention have been focused on the charge-ordered 
compositions of Pr$_{1-x}$Ca$_x$MnO$_3$ due to their importance as 
possible prototypes for the electronic phase separation (PS) 
models \cite{deac,fisher,kajimoto,nagapriya,moreo1,moreo2,yunoki,sarma1,radaelli} 
proposed to explain the phenomenon of colossal magnetoresistance (CMR). 
The PS models are qualitatively different from the double-exchange model \cite{zener,anderson,gennes} or those based on strong Jahn-Teller 
polarons \cite{millis1,millis2}. According to the PS model the ground state 
of CMR materials is comprised of coexisting nanosize clusters of metallic 
ferromagnetic and insulating antiferromagnetic nature \cite{moreo2}. The 
insulator-metal transition in this scenario is through current 
percolation. Though there have been a number of experimental studies 
showing the existence of phase separation, their size varies from nano- 
to mesoscopic scales \cite{sarma1,radaelli}. Radaelli et al. have shown the 
origin of mesoscopic phase separation to be the intergranular 
strain \cite{radaelli} rather than the electronic nature as in PS models. 
Nanosized stripes of a ferromagnetic phase \cite{simon} were reported in 
Pr$_{0.67}$Ca$_{0.33}$MnO$_3$. This compound also attracted much 
attention earlier due to the existence of a nearly degenerate 
ferromagnetic metallic state and a charge-ordered antiferromagnetic 
insulating state with a field-induced phase transition possible between 
them \cite{miyano}. With slight variations in the Ca doping the 
Pr$_{1-x}$Ca$_x$MnO$_3$ system turns ferromagnetic ($x$=0.2) or 
antiferromagnetic ($x$=0.4) at low temperatures \cite{wollam,simon}. The 
composition $x$=0.33 shows a coexistence of ferromagnetic and 
antiferromagnetic phases \cite{jirak}. This makes the charge- orbital-ordered 
Pr$_{0.67}$Ca$_{0.33}$MnO$_3$ a prototype for the PS scenario.

One of the requisites for the existence of an electronic phase separation 
in manganites is a strong-coupling interaction affecting the hopping of 
the itinerant $e$$_g$ electrons. The PS is expected to occur \cite{moreo1} 
when $J$$_H$/$t$ $\gg$ 1, where $J$$_H$ is the Hund's coupling contribution 
between the localized $t$$_{2g}$ and the $e$$_g$ electrons and $t$ is the 
hopping amplitude of the e$_g$ electrons. The PS also favors a strong 
electron-phonon coupling, like the influence of a strong Jahn-Teller (JT) 
polaron arising from the $Q$$_2$ and $Q$$_3$ JT modes. Essentially, one 
expects a strong localization of charge carriers to accompany the 
electronic separation of phases. Apart from charge, the orbital degrees of 
freedom also play an important role in this scenario \cite{moreo1}. The 
itinerant electron hopping term is strongly influenced by the symmetry of 
the orbitals ($d_{x^2-y^2}$ and $d_{3z^2-r^2}$) hybridized with the O 2$p$ 
orbitals of the MnO$_6$ octahedra. In comparison with the most popular CMR 
material La$_{1-x}$Sr$_x$MnO$_3$, the charge-ordered Pr system has an 
inherently reduced $e$$_g$ bandwidth $W$ due to the smaller ionic radius of 
Pr, which also enhances the tendency for carrier localization. All these 
charge and orbital interactions are reflected in the near-$E$$_F$ electronic 
structure of these materials and could be probed using electron 
spectroscopic techniques. Valence band photoemission and O $K$ x-ray 
absorption are two such proven tools sensitive to the changes in the low-energy states crucial to these interactions. Although, there are many 
reports on the near-$E$$_F$ electronic structure of other CMR compounds 
using these 
techniques \cite{sarma2,park1,park2,saitoh1,saitoh2,pal,dessau1,dessau2,mannella,toulemonde} 
only a few studies have been reported on their charge-ordered 
compositions \cite{sarma1,park3}. One of the key energy terms these 
spectroscopies could show earlier was the charge transfer energy $E$$_{CT}$ 
which is intimately related to the electron-electron and electron-lattice 
interactions \cite{park1}.

In this study we have used ultraviolet photoelectron spectroscopy and x-ray absorption spectroscopy (XAS) in order to probe the electronic 
structure of the occupied and unoccupied states on a well-characterized, 
high-quality single crystal of Pr$_{0.67}$Ca$_{0.33}$MnO$_3$. Another part 
of this single crystal had earlier been used for a detailed neutron 
scattering experiment that indicated the possible existence of a phase 
separation of ferromagnetic and antiferromagnetic stripes \cite{simon}. In 
the present study we have analyzed the temperature-dependent changes in 
the near-$E$$_F$ electron energy states from the perspective of a phase 
separation.

\section{EXPERIMENT}

The single-crystal sample of Pr$_{0.67}$Ca$_{0.33}$MnO$_3$ was grown by 
the floating zone method in a mirror furnace. The compositional 
homogeneity of the crystal was confirmed using energy-dispersive 
spectroscopic analysis. Magnetization and transport measurements on 
this crystal showed the transition temperatures $T$$_c$ and $T$$_N$ to be 100 
and 110 K, respectively. Details of the sample preparation, magnetization, 
and transport measurements and structural studies are published 
elsewhere \cite{simon,silvina}. Angle-integrated ultraviolet photoemission 
measurements were performed using an Omicron ${\mu}$-metal UHV system equipped 
with a high-intensity vacuum-ultraviolet source (HIS 13) and a hemispherical electron energy analyzer (EA 125 HR). At the He $I$ (21.2 eV) line, the photon flux 
was of the order of 10$^{16}$ photons/s/sr with a beam spot of 
2.5 mm diameter. Fermi energies for all measurements were calibrated 
using a freshly evaporated Ag film on a sample holder. The total energy 
resolution, estimated from the width of the Fermi edge, was about 80 meV. 
The single-crystalline samples were repeatedly scraped using a 
diamond file inside the chamber with a base vacuum of 
$\sim$ 1.0 $\times$ 10$^{-10}$ mbar. Scraping was repeated until negligible intensity was found for the bump around 9.5 eV, which is a signature of surface 
contamination \cite{sarma2}. For the temperature-dependent measurements, 
the sample was cooled by pumping liquid nitrogen through the sample 
manipulator fitted with a cryostat. Sample temperatures were measured 
using a silicon diode sensor touching the bottom of the sample holder. XAS 
measurements were performed using the BEAR (Ref. 31) and 
BACH (Ref. 32) beamlines associated with ELETTRA at Trieste, Italy. 
At the BEAR beamline we used monochromatized radiation from a bending 
magnet in order to record the O $K$  edge spectra at room temperature and 150 
K in fluorescence detection mode on a freshly scraped surface of the 
single crystal. The energy resolution was around 0.2 eV in the case of 
these two spectra. The O $K$ edge spectra at 95 K was recorded at the BACH 
beamline, using the total electron yield mode. Before the 
measurements, the sample surface was scraped inside the UHV chamber 
($\sim$ 1.0 $\times$ 10$^{-10}$ mbar) using a diamond file. At this beamline we 
used radiation from an undulator, monochromatized using a spherical 
grating. The resolution at the O $K$edge for this measurement was better 
than 0.1 eV.

\section{RESULTS AND DISCUSSION}

The angle-integrated valence band photoemission spectra of 
Pr$_{0.67}$Ca$_{0.33}$MnO$_3$ taken at different temperatures below and 
above $T$$_c$ are shown in Fig. 1. Intensities of all the spectra were 
normalized and shifted along the $y$ axis by a constant for clarity. The 
spectra, dominated by the states due to the Mn 3$d$-O 2$p$ hybridized 
orbitals, look similar to those reported earlier on the 
La$_{1-x}$Sr$_x$MnO$_3$ system \cite{sarma2,park1,park2,saitoh1,saitoh2}. 
The origin of the two prominent features, one at $\sim$ -3.5 eV (marked $B$) 
and another at $\sim$ -5.6 eV (marked $C$) below $E$$_F$, are now well 
known from earlier experiments and band structure 
calculations \cite{saitoh2,pal,kurmaev} on similar systems. While the 
feature at -3.5 eV is mainly due to the $t$$_{2g\uparrow}$ states of the 
MnO$_6$ octahedra, the -5.6 eV subband has contributions from both 
$t$$_{2g}$ and $e$$_g$ states. More important contributors to the properties 
of these systems are the states nearer to $E$$_F$, which appear as a 
tail at $\sim$ -1.2 eV (marked $A$) from the chemical potential. The Intensity 
of this feature is quite small compared to $B$ and $C$ and is a signature of 
the insulating nature of this material. Earlier photoemission experiments 
on La$_{1-x}$Sr$_x$MnO$_3$ also have shown that the intensity of this tail 
feature $A$ is quite small \cite{sarma1,park1,park2,saitoh1,saitoh2}. 
The presence of the feature $A$ is clear from Fig. 2, where we have shown 
the near-$E$$_F$ region of the valence band spectra taken with a higher 
resolution. Here, the spectra taken at different temperatures have been 
compared with those at room temperature. The figure also shows the 
difference spectra obtained by subtracting the room-temperature spectra 
from spectra taken at low temperatures. The feature in the difference 
spectra corresponds to $A$, which originates from the $e$$_{g\uparrow}$ 
states \cite{saitoh1}. In order to estimate the relative changes in all the 
three features ($A$, $B$, and $C$) due to temperature, we have carefully fitted 
the whole valence band spectrum with three components corresponding to the 
three subbands contributing to this region. We used a $\chi$$^{2}$ 
iterative program for fitting the different spectra with Lorentzian line 
shapes for $A$, $B$, and $C$. Except for the area of the peaks, the energy positions, widths, background, and all other parameters of all the peaks were kept the same for different temperatures. The positions and full widths at half maximum (FWHMs) of the three peaks are given in the caption of Fig. 3. The fitted spectra for 77 K (below $T$$_c$) and 300 K (above $T$$_c$) are shown on the left sides of 
Figs. 5(a) and 5(b) below. One can see from Figs. 3(a)-3(c) and that the area of all the peaks keeps increasing as we go down in temperature $T$$_c$, but below this temperature the intensities of both peaks $C$ and $B$ decrease. 
On the other hand, the intensity of $A$ does not show any decrease across 
the transition [Fig. 3(a)].

In a reasonable approximation, the near-$E$$_F$ region of the O $K$ x-ray 
absorption spectra could well represent the density of unoccupied states 
in many of the transition metal oxide compounds \cite{park1,degroot}. 
In order to probe the electronic structure of the unoccupied states we 
have performed XAS on our Pr$_{0.67}$Ca$_{0.33}$MnO$_3$ single crystal. 
The O $K$ edge XAS spectra taken at room temperature and 150 and 95 K (below 
$T$$_c$), shown in Fig. 4, have been normalized in intensity all along the 
region starting from 550 eV. The preedge feature in the O $K$ spectra 
(centered around 529.5 eV) is due to the strong hybridization between Mn 
3$d$ and O 2$p$ orbitals. We will analyze this structure in detail in the 
later part of this paper. The broad feature around 536.5 eV is due to the 
bands from hybridized Pr 5$d$ and Ca 3$d$ orbitals, while the structure above 
540 eV is due to states like Mn 4$sp$ and Pr 6$sp$, etc. These states are known 
to contribute least to the near-$E$$_F$ electronic structure of these 
transition metal-oxide compounds. The assignments of the features 
mentioned above are consistent with the band structure calculations on 
similar systems \cite{degroot,kurata}.

The preedge feature in the O $K$ edge spectra carries a substantial amount 
of physics involved in the properties of these materials. It has earlier 
been shown that the prepeak in the O $K$ edge spectra of different CMR 
materials consists of two lines, one of which appears as a shoulder on the 
low-energy side of the other \cite{park2,toulemonde,dessau1,mannella}. 
The intensity of this shoulder was found to increase as the material goes 
across the insulator-metal transition \cite{toulemonde,mannella}. Since our 
Pr$_{0.67}$Ca$_{0.33}$MnO$_3$ sample is an insulator at all temperatures 
the prepeak in our O $K$ edge spectra (Fig. 4) does not show any 
splitting, though the presence of the shoulder is visible as an asymmetry 
on the low-energy side of this peak. This shoulder feature is due to 
the first available unoccupied states and in XAS corresponds to the 
addition of an electron to the $e$$_{g\uparrow}$ state of the crystal-field-split
MnO$_6$ octahedra. The main feature in the prepeak arises from the 
$t$$_{2g\downarrow}$ states \cite{park2}. For a quantitative estimate of the 
temperature-dependent changes in the intensities of the two features we 
have fitted the prepeak using two components of Lorentzian line shapes. 
Here also the data fitting was done using $\chi$$^{2}$ iterative method, 
keeping the energy positions and widths the same for all temperatures. Results 
of the data fit are shown in Table I. Fitted spectra for temperatures 
above (300 K) and below $T$$c$ (95 K) are shown on the right side of 
Figs 5(a) and 5(b). Though very small, the intensity of the feature $A$$^{\prime}$ shows 
a slight increase below $T$$_c$. The best fit gives a 0.6$\pm$0.1 eV 
difference between the energy positions of $A$$^{\prime}$ and $B$$^{\prime}$. Following a different assignment of origin to the prepeak features Dessau $et$ $al$. have interpreted this energy difference as a measure of the Jahn-Teller 
distortion (4$E$$_{JT}$) \cite{dessau2}.

Having shown the temperature-dependent changes in the UPS and XAS results 
in Figs. 2 and 4, we present a combined picture of them in order to 
derive some insights into the charge transfer energy involved in the 
properties of the Pr$_{0.67}$Ca$_{0.33}$MnO$_3$ sample. The combined 
spectra of the valence band from photoemission and the preedge region of 
the O $K$ edge XAS, presented in Figs 5(a) and 5(b) give the density of both 
occupied and unoccupied states above and below the Curie temperature 
($T$$_c$). A schematic of these energy levels is shown in Fig. 6. The 
hole or electron doping (value of $x$) causes the symmetry of the last occupied and first unoccupied bands to be $e_{g{z^2 - r^2}\uparrow}$. In Fig. 5 these 
states are marked by $A$ and $A$$^{\prime}$. In order to scale the left and right sides of $E$$_F$, the O $K$ edge spectra have been shifted such that the energy 
of the first unoccupied state (1.4 eV from $E$$_F$) found from inverse 
photoemission \cite{chainani} coincides with that of $A$$^{\prime}$. We have used 
integral backgrounds for both sides of $E$$_F$ and, as mentioned earlier, 
all the fitting parameters were kept the same including the energy 
positions for the high and low temperatures. The important parameter which 
can be derived from this combined spectrum is the charge transfer energy 
$E$$_{CT}$, which is the energy required for an $e$$_g$ electron to hop 
between the Mn$^{3+}$($t$$^3$$_{2g}$$e$$^1$$_g$) and Mn$^{4+}$($t$$^3$$_{2g}$) 
sites. In our spectra (Fig. 5) $E$$_{CT}$ is the energy difference 
between $A$ and $A$$^{\prime}$ which corresponds to the last occupied and first 
unoccupied states. The value of $E$$_{CT}$ is 2.6$\pm$0.1 eV from our 
spectra, which is higher than the value (1.5$\pm$0.4 eV) determined {\it indirectly} by Park $et$ $al$. for the La$_{1-x}$Ca$_x$MnO$_3$ system \cite{park1}. The larger value of $E$$_{CT}$ shows the existence of a stronger charge localization 
in the charge- or orbital-ordered Pr$_{0.67}$Ca$_{0.33}$MnO$_3$ compound.

Pr$_{0.67}$Ca$_{0.33}$MnO$_3$ shows a transition to a charge-ordered state 
below 220 K (T$_{co}$) \cite{silvina,christine,hardy}. Neutron diffraction 
studies \cite{simon} on another part of this single crystal have shown that 
this charge ordering turns into an antiferromagnetically structured 
pseudo-CE type charge or orbital ordering below 110 K. Also, the coexistence 
of ferromagnetic and antiferromagnetic phases below this transition 
temperature has been shown on this single crystal. In the pseudo-CE-type 
ordering the Mn 3$d_{z^2-r^2}$ orbitals, where the $e$$_g$ state is 
occupied, are aligned with the O 2$p$ orbitals. Such an alignment can 
increase the hybridization between Mn 3$d_{z^2-r^2}$ and O 2$p$. 
Furthermore, a simultaneous decrease in the hybridization strength could 
also be expected for the corelike in-plane $t$$_{2g}$ and O 2$p$ states. 
The results of the curve fitting (Fig. 3) of our valence band spectra 
reflects these changes in hybridization with temperature. The charge 
ordering following the decrease in temperature causes an increase in the O 
2$p$ contribution to both the $t$$_{2g}$ and $e$$_g$ spin-up subbands in the 
valence region and hence the intensities of $A$, $B$, and $C$ go up. As mentioned 
earlier, the pseudo-CE-type charge or orbital ordering below 110 K, results 
not only in a stronger hybridization of O 2$p$ with the $e_{g{z^2 - 
r^2}\uparrow}$ states compared to that with the $t$$_{2g}$ states but a 
simultaneous decrease in the latter also. This is reflected in the 
increase in intensity of $A$ and decrease in the intensities of both $C$ and $B$ 
below 110 K. Though in Pr$_{0.67}$Ca$_{0.33}$MnO$_3$ there is no 
temperature-dependent insulator-metal transition, the change in 
intensities of these features across $T$$_c$ appears similar to the shifting 
of spectral weight found in the La$_{1-x}$Sr$_x$MnO$_3$ 
system \cite{sarma1}.

The value of $E$$_{CT}$ indicating a strong charge localization, found in 
our Pr$_{0.67}$Ca$_{0.33}$MnO$_3$ has significant implications for the 
models proposed to explain the CMR effect in manganites. Charge 
localization can result from strong-coupling interactions of the $e$$_g$ 
electron with the corelike $t$$_{2g}$ or the $Q$$_2$ and $Q$$_3$ Jahn-Teller 
modes. Both these interactions influence the electron hopping term ($t$). 
The former interaction, ferromagnetic in nature, is the Hund's coupling 
$J$$_H$. A large value of $J$$_H$ favors the electronic separation of phases. 
A strong electron-lattice coupling results from the cooperative 
Jahn-Teller distortions of the MnO$_6$ octahedra. These distortions, 
particularly the $Q$$_2$ and $Q$$_3$ vibrational modes of the oxygen ions are 
stronger in the case of Pr-containing manganites compared to La-containing 
ones. Park $et$ $al$. \cite{park1} have reported a smaller charge transfer 
energy ($E$$_{CT}$ = 1.5 eV) in the case of La$_{1-x}$Ca$_x$MnO$_3$ and have 
attributed it to small polarons (Anderson localization) induced from 
the ionic size difference between Mn$^{3+}$ and Mn$^{4+}$. A small-polaronic
model may not be able to explain the high value of $E$$_{CT}$ 
found in Pr$_{0.67}$Ca$_{0.33}$MnO$_3$. On the other hand, a strong charge 
localization is expected in the case of the PS model in which the ground state 
is described as a mixture of ferromagnetic and antferromagnetic regions. It is also possible that 
this large charge localization is due to the Zener polaron proposed by 
Aladine $et$ $al$ \cite{aladine}. In the Zener polaron picture, the Mn ions of 
adjacent MnO$_6$ octahedra form a dimer with variations in the Mn-O-Mn 
bond angle. The pseudo-CE-type charge ordering in Pr$_{1-x}$Ca$_{x}$MnO$_3$ 
favors such regular distortions in the MnO$_6$ octahedra. Further studies using electron spectroscopic techniques on samples in the vicinity of the charge-ordered compositions may be able to differentiate between these possible driving mechanisms behind the strong charge localization.

\section{CONCLUSIONS} 

Using valence band photoemission and O $K$ edge x-ray absorption, we have 
probed the electronic structure of Pr$_{0.67}$Ca$_{0.33}$MnO$_3$, which is 
regarded as a prototype for the electronic phase separation models in CMR 
systems. With decrease in temperature the O 2$p$ contributions to the 
$t$$_{2g}$ and $e$$_g$ spin-up states in the valence band were found to 
increase until $T$$_c$. Below $T$$_c$, the density of states with $e$$_g$ 
spin-up symmetry increased while those with $t$$_{2g}$ symmetry decreased, 
possibly due to the change in the orbital degrees of freedom associated 
with the Mn 3$d$ - O 2$p$ hybridization in the pseudo-CE-type charge or orbital 
ordering. These changes in the density of states could well be connected 
to the electronic phase separation reported earlier. Our quantitative 
estimate of the charge transfer energy ($E$$_{CT}$) is 2.6$\pm$0.1 eV, 
which is large compared to the earlier reported values in other CMR 
systems. Such a large charge transfer energy may support the phase 
separation model.  

\section*{ACKNOWLEDGMENTS}
The authors would like to thank the staff at the BEAR and BACH beamlines 
of Elettra Sincrotrone Trieste, Italy for the XAS measurements and DST, 
India for financial support.

\begin{widetext}
\newpage
\begin{table}[h]
\begin{center}
\caption{$\chi$$^{2}$ iterative fitting parameters for the preedge peak in 
O $K$ XAS. The intensities of $B$$^{\prime}$ and $A$$^{\prime}$ at low temperature are normalized with respect to the total intensity of the prepeak at room temperature. We have used an integral background, which was kept the same for all the spectra. The energy positions and FWHMs were determined by finding the best 
fit common to the three spectra by a $\chi$$^{2}$ iterative program. The 
final fit for all the spectra from different temperatures were obtained 
with the same energy positions and FWHMs.\\
$~~$\\}
\begin{tabular}{|c|c|c|c|c|c|c|}
\hline
Temperature (K) & \multicolumn{6}{c|} {O $K$ edge x-ray absorption spectra} \\
\cline{2-7}
&\multicolumn{3}{c|}{$B$$^{\prime}$}&\multicolumn{3}{c|}{$A$$^{\prime}$}\\
\hline
& Position & FWHM & Normalized & Position & FWHM & Normalized \\
95&(eV)&(eV)&area&(eV)&(eV)&area\\
\cline{2-7}
&2.11&1.52&0.72&1.48&0.97&0.29\\
\hline
150&2.11&1.52&0.72&1.48&0.97&0.26\\
\hline
300&2.12&1.52&0.73&1.47&0.97&0.26\\
\hline
\end{tabular}
\end{center}
\end{table}

\newpage

\begin{figure}
\includegraphics[width=6.0in]{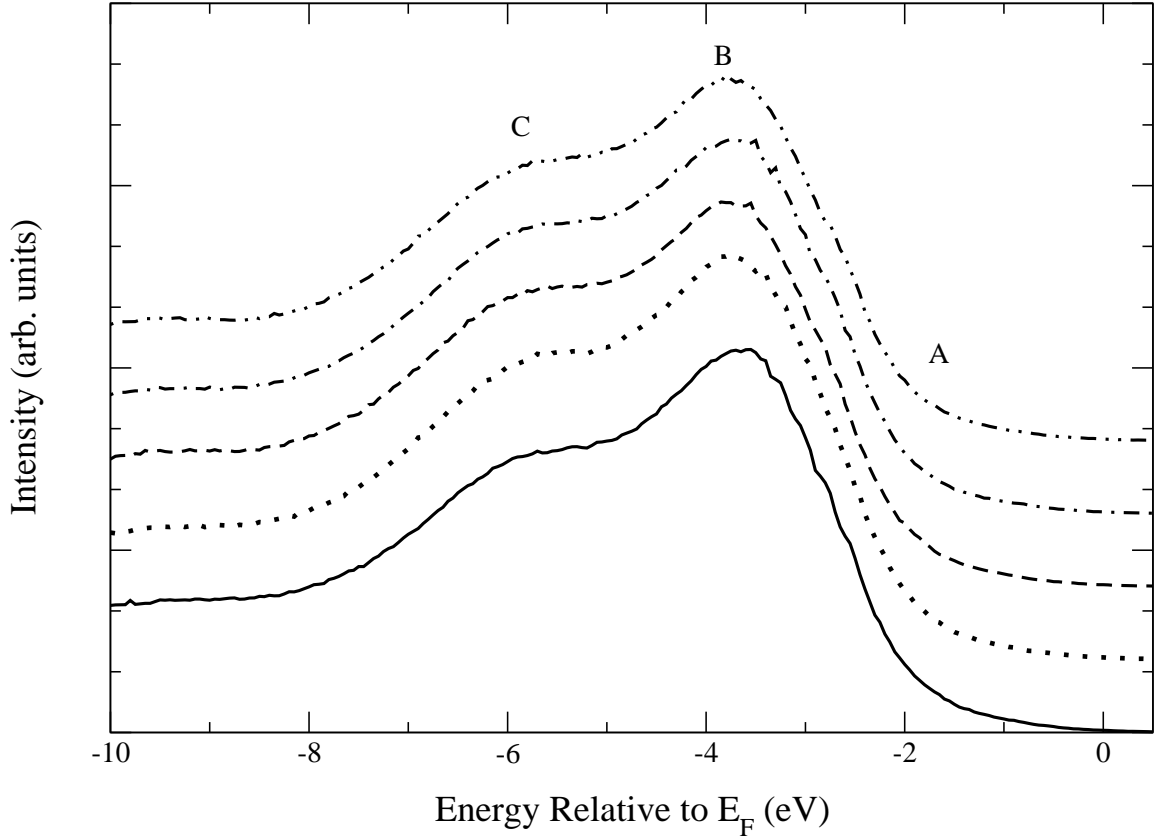}
\caption{\label{figure_1} Valence band photoemission spectra of 
Pr$_{0.67}$Ca$_{0.33}$MnO$_3$ taken using He $I$ photon energy (21.2 eV) at 
77 (solid line), 110 (dotted), 150 (dashed), 220 (dot-dashed), and 300 K 
(double-dot dashed). All the spectra have been normalized and shifted along y-axis by a constant for clarity. The subbands around -1.2 ($e$$_{g\uparrow}$), 
-3.5 ($t$$_{2g\uparrow}$), and -5.6 eV ($e$$_{g\uparrow}$ + 
$t$$_{2g\uparrow}$) are marked as $A$, $B$, and $C$ respectively.} 
\end{figure}

\newpage   
\begin{figure}
\includegraphics[width=6.0in]{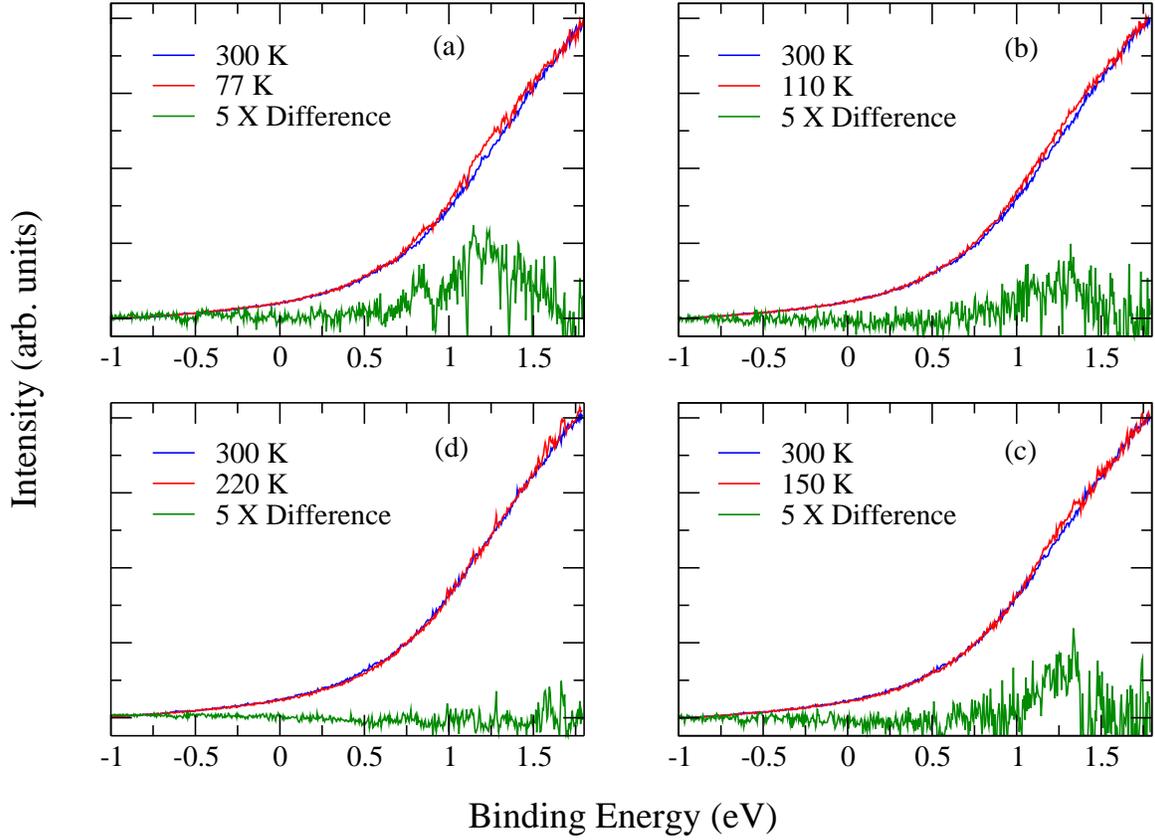}
\caption{\label{figure_2} (color online) High-resolution photoemission spectra of the near-$E$$_F$ region of the valence band of Pr$_{0.67}$Ca$_{0.33}$MnO$_3$. In 
(a) the spectrum taken below $T$$_c$ (red) is compared with the normal 
state (300 K) spectrum (blue). The difference spectrum (green) obtained by 
subtracting the spectra at 300 K from 77 K and multiplied by 5 is also 
shown in the panel. Similarly, in (b), (c), and (d) the near-$E$$_F$ 
spectra taken at 110, 150, and 220 K are compared with that taken at 300 
K. The feature in the difference spectra corresponds to the tail feature $A$ 
($e$$_{g\uparrow}$) in Fig. 1.}
\end{figure}

\newpage
\begin{figure}
\includegraphics[width=6.0in]{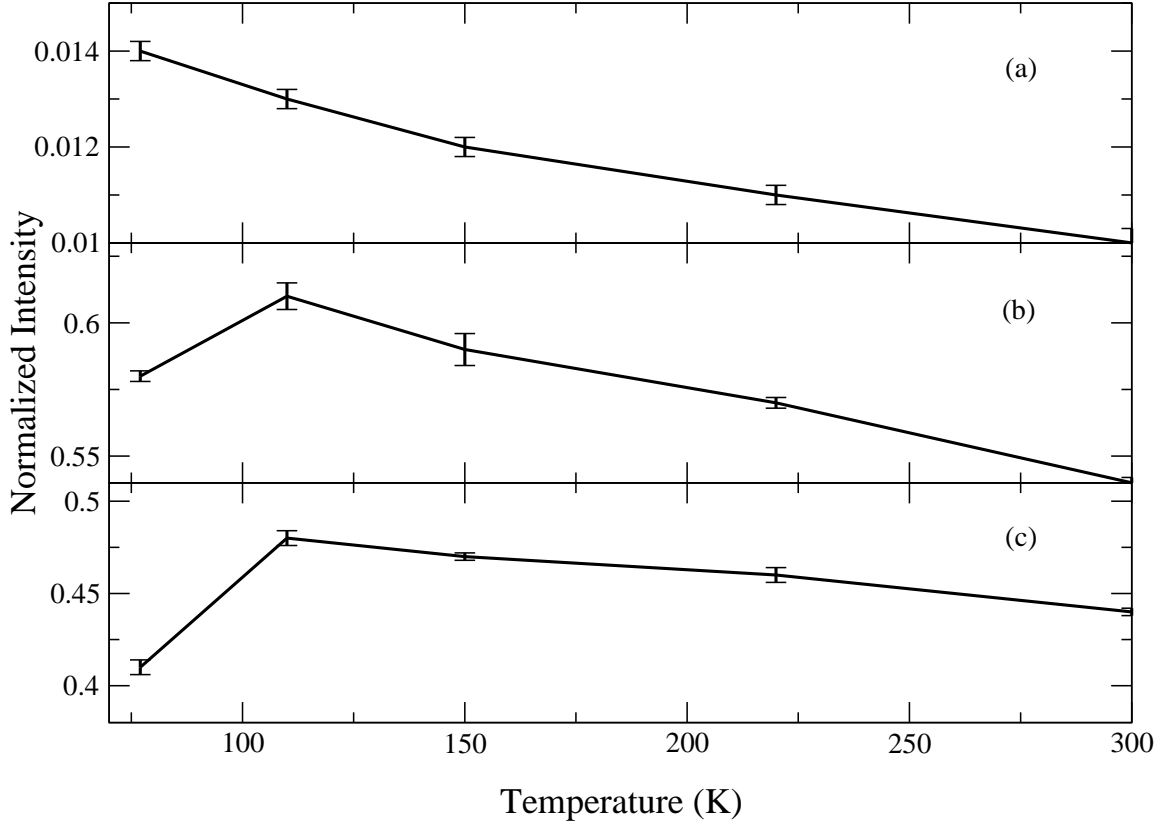}
\caption{\label{figure_3} Temperature dependence of the area of the three 
valence band features obtained from fitting the spectra with Lorentzian 
line shapes using a $\chi$$^{2}$ iterative program. We have used an 
integral background, which was kept the same for all the spectra. The 
energy positions and FWHMs were determined by finding the best fit common
to all the spectra by the iterative program. The final fit for all 
spectra at different temperatures were obtained with the same energy 
positions and FWHMs. (a), (b), and (c) correspond to the features $A$, $B$, 
and $C$ positioned at -1.19, -3.49, and -5.63 eV with FWHMs 2.56, 
1.93, and 1.37 eV, respectively.} 
\end{figure}

\newpage
\begin{figure}
\includegraphics[width=6.0in]{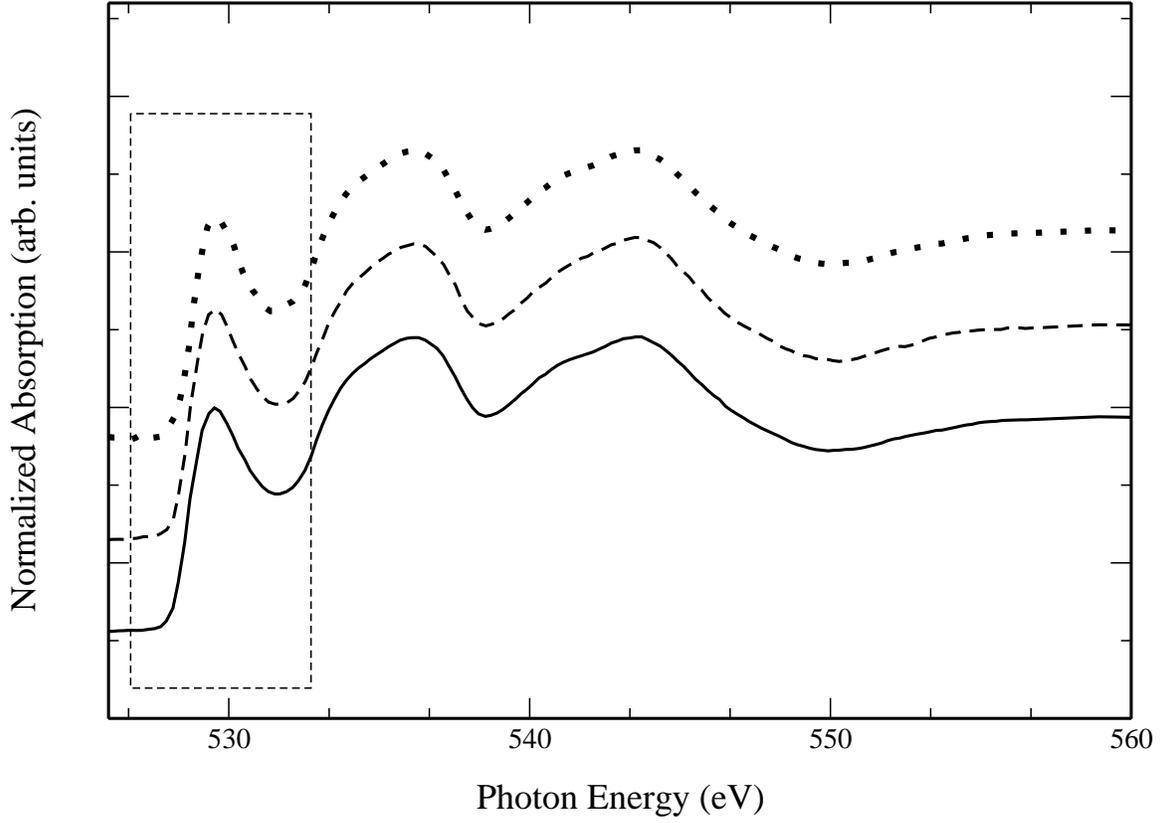}
\caption{\label{figure_4}O $K$ edge x-ray absorption spectra of 
Pr$_{0.67}$Ca$_{0.33}$MnO$_3$ taken at 300 (solid line), 150 (dashed 
line), and 95 K (dotted). The preedge feature centered around 529.5 eV 
(marked by a box) is where most interest lies. Since this feature consists 
of two peaks (a main line and a shoulder on the low-energy side), this 
part of the spectrum was fitted with two components of Lorentzian line 
shapes. Results of the curve fit are given in Table I.} 
\end{figure}

\newpage
\begin{figure}
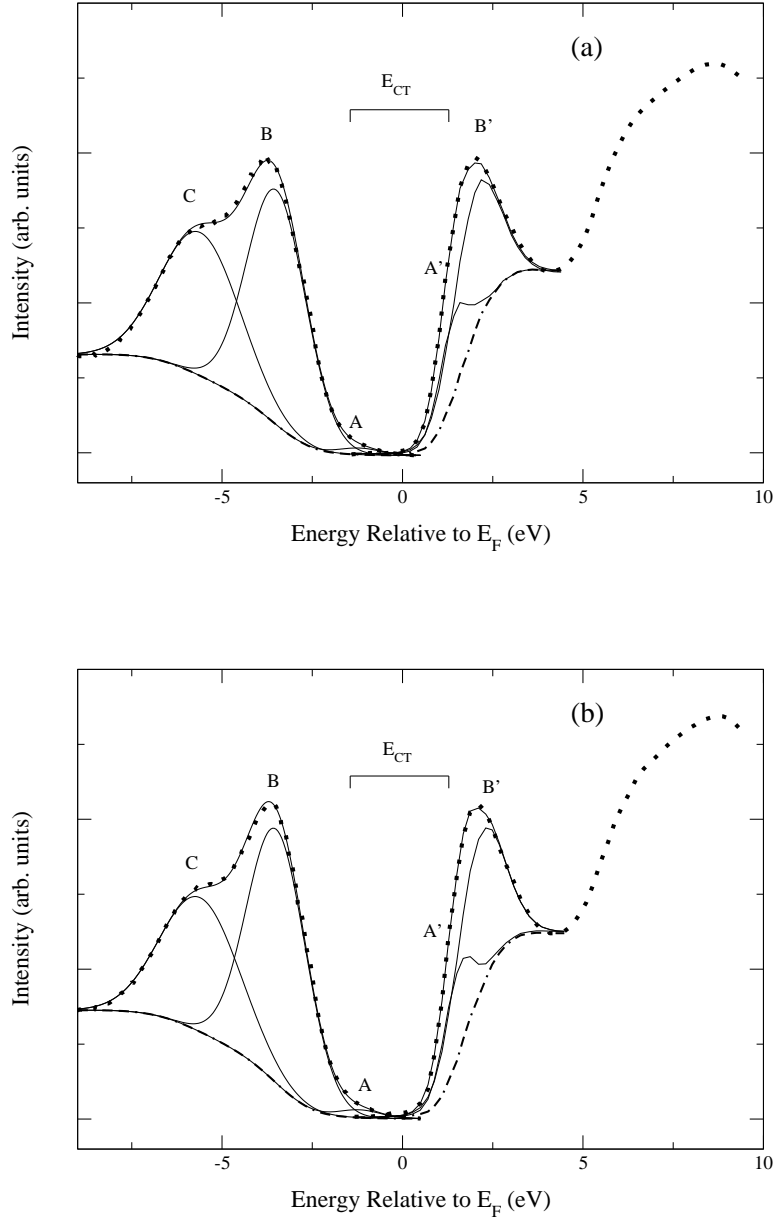

\includegraphics[width=4.0in]{figure5a.eps}\\
\vspace{1.5cm}
\includegraphics[width=4.0in]{figure5b.eps}
\caption{\label{figure_5} (a) Combined spectra from valence band 
photoemission (occupied energy states) and prepeak in O $K$ edge x-ray 
absorption spectra (unoccupied energy states) of 
Pr$_{0.67}$Ca$_{0.33}$MnO$_3$ taken at room temperature (above $T$$_c$). The 
three features corresponding to the subbands in the valence region are 
marked $A$, $B$, and $C$.The two peaks of the fitted O $K$ edge prepeak are 
marked $A$$^{\prime}$ and $B$$^{\prime}$. We used integral background for both sides of $E$$_F$. 
Details of the fitting are mentioned in the text. (b) Combined spectra of 
Pr$_{0.67}$Ca$_{0.33}$MnO$_3$ below $T$$_c$. The valence band spectrum was 
taken at 77 K and O $K$ edge XAS was taken at 95 K.}
\end{figure}

\newpage
\begin{figure}
\includegraphics[width=6.0in]{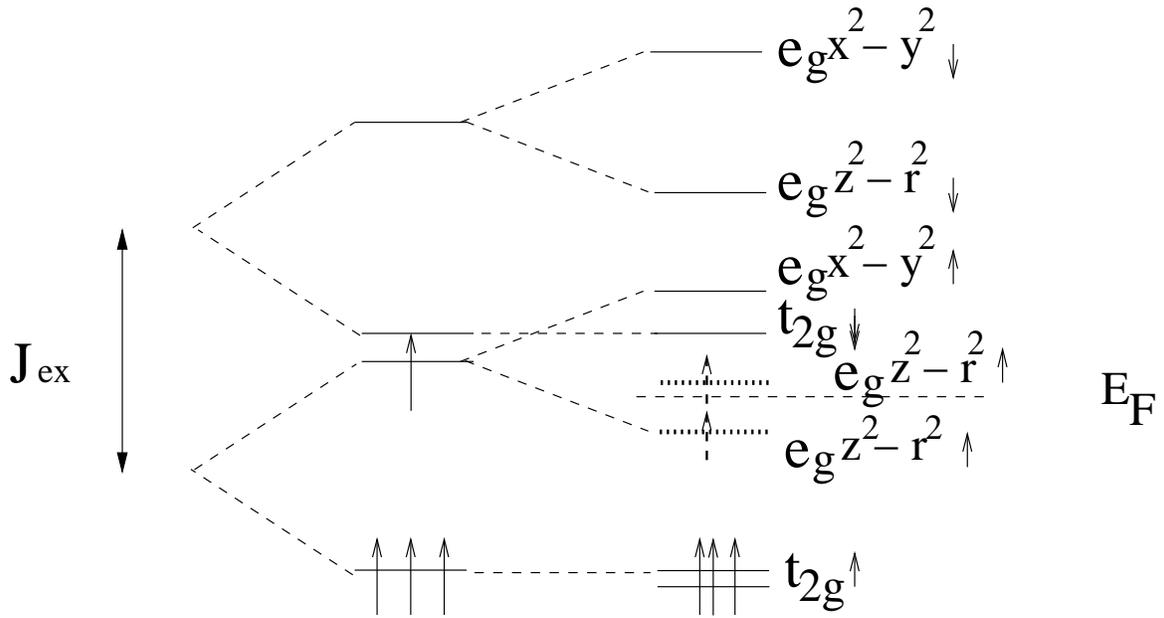}
\caption{\label{figure_6} Schematic diagram of the near-$E$$_F$ energy 
levels of Pr$_{0.67}$Ca$_{0.33}$MnO$_3$ in the occupied and unoccupied 
parts. The diagram is not drawn to scale. Some of the $e$$_{g\uparrow}$ 
states are occupied (at the Mn$^{3+}$ sites) and some are unoccupied 
(at the Mn$^{4+}$ sites). Hence the last occupied and first unoccupied 
states are marked with dotted lines.}
\end{figure}
\end{widetext}

\end{document}